\documentclass[prd,aps, nofootinbib,10 pt]{revtex4}
\usepackage{amssymb}
\usepackage{amsmath,graphicx,color,epsfig}
\usepackage{subfig}
\usepackage{epsfig}
\usepackage{pstricks}
\usepackage{float}

\begin{document}

\title{Spin-orbit splittings in heavy-light mesons and Dirac equation}
\author{Riazuddin}
\email{riazuddin@ncp.edu.pk}
\affiliation{National Centre for Physics,\\
Quaid-i-Azam University Campus, Islamabad 45320, Pakistan }
\author{Sidra Shafiq}
\affiliation{Cenre for Advance Mathhematics and Physics,\\
National
University of Science and Technology, Islamabad, Pakistan}
\date{\today }

\begin{abstract}
The spin-orbit splitting in heavy-light mesons is seen to be
suppressed experimentally. It is shown that it can be understood
qualitatively in the frame work of Dirac theory. An alternative
derivation of a relativistic dynamical symmetry for the Dirac
Hamiltonian, which suppresses spin orbit splitting, is also given.
However it is shown that such a symmetry is not needed since the
spin-orbit splitting in Dirac theory with Coulomb like potential (as
is the case for the one gluon exchange potential in pQCD) is small
anyway.
\end{abstract}

\maketitle

\section{Introduction}

In the mass spectroscopy for heavy-light mesons as $q\overline{Q}$
or $Q\overline{q}$ bound states, one has both hyperfine splitting
and spin-orbit splitting, which need to be understood in quantum
chromodynamics (QCD). Writing a pseutoscalar(vector) heavy meson as
$P_{q}(P_{q}^{\ast })$, $P=D$ or $B$, $q=u$, $d$ or $s$, the
hyperfine splittings in MeV experimentally are [1]
\begin{eqnarray}
M_{D_{d}^{\ast }}-M_{D_{d}^{\ast }} &=&140.64\pm 0.10\simeq
M_{D_{s}^{\ast
}}-M_{D_{s}^{\ast }}=143.8\pm 0.04  \notag \\
M_{B_{d}^{\ast }}-M_{B_{d}^{\ast }} &=&45.78\pm 0.35\simeq
M_{B_{s}^{\ast }}-M_{B_{s}^{\ast }}=46.5\pm 1.2  \label{1}
\end{eqnarray}%
On the other hand the spin orbit splittings seem to be suppressed
(see below).

We can write the effective Hamiltonian for a bound hadron containing
one heavy quark(antiquark) $Q(\overline{Q})$ as
\begin{equation}
H=H_{q}+H_{Q}  \label{2}
\end{equation}%
where $H_{Q}$\ takes care of the residual momentum of the heavy
quark(antiquark) and $H_{q}$ represent the motion of the light
antiquark(quark) in a fixed potential provided by the heavy
quark(antiquark). In heavy quark effective theory[2] (HQET), $H_{Q}$
contains a term
$\overrightarrow{\sigma}_Q.\overrightarrow{B}_c/2m_Q$ which gives
rise to color magnetic moment interaction of type
$\overrightarrow{\mu}_{q}.\overrightarrow{\mu}_{Q}$ which induces
the conventional form of the Fermi-Breit potential
\begin{equation}
\frac{8\pi }{3}\alpha _{V}(\mu )\frac{\sigma _{Q}.\sigma _{q}}{4m_{Q}m_{q}}%
\delta ^{3}(r)  \label{3}
\end{equation}%
where $\alpha _{V}(\mu )=\frac{4}{3}\alpha _{G}(\mu )$, $\alpha
_{G}$\ is the pQCD running coupling constant which depends on the
energy scale $\mu $, in our case mass of the heavy quark. Here $m_Q$
and $m_q$ are effective constituent quark masses. The hyperfine
splitting can be understood [3] in terms of the above term in
$H_{Q}$, which shows that such splittings decrease with the
increasing mass of $Q$ both because of $m_{Q}$\ in the denominator
and decrease of $\alpha _{s}(\mu )$\ due to asymptotic freedom
property of pQCD.

Since in HQET\ the spin of the heavy quark is decoupled it is
natural to combine
$\overrightarrow{j}=\overrightarrow{L}+\overrightarrow{S}_{q}$, the
angular momentum of light degrees of freedom, with
$\overrightarrow{S}_{Q}$\ to give
$\overrightarrow{J}=\overrightarrow{j}+\overrightarrow{S}_{Q}$ for
the bound $Q\overline{q}$ system. Thus one can have the following
multiplets
\begin{eqnarray}
l &=&0\text{ \ \ }\left[ P^{\ast }(1^{-}),P(0^{-})\right]
_{j=\frac{1}{2}}
\notag \\
l &=&1\text{ \ \ }\left[ P_{2}^{\ast }(2^{+}),P_{1}(1^{+})\right] _{j=\frac{3%
}{2}}  \notag \\
&&\left[ P_{1}^{\ast }(1^{+}),P_{0}(0^{+})\right] _{j=\frac{1}{2}}
\label{4}
\end{eqnarray}
where $P$ is $D$ or $B$ and $J^{P}$ gives the total angular momentum
and parity quantum numbers. The splitting between $j=\frac{3}{2}$
and $j=\frac{1}{2}$ for $l=1$ is due to the spin-orbit coupling
$\overrightarrow{L}.\overrightarrow{S}_q$ while the hyperfine
splitting between 2 members of each multiplet arise from the
Fermi-Breit term as mentioned earlier. The spin-orbit splitting (in
MeV)[1] for $D$ mesons, between $D_{2}^{\ast}(2^{+}):2462.8\pm1.0$
and $D_{1}^{\ast }(1^{+}):2422.3\pm 0.6$MeV is 40, for the $B$
mesons, between $B_{2}^{\ast }(2^{+}):5743.9\pm 5.0$ and
$B_{1}^{\ast }:5723.4\pm 2.0$ is 21, for
$B^{2*}_{s_{2}}(2^{+}):5839.7\pm 0.6$ and
$B_{s_{1}}(1^{+}):5829.4\pm0.7$ is 11. Thus these splittings are
suppressed. A measure of this suppression is the parameter [4]
\begin{eqnarray}
r=\frac{p_{3/2}-p_{1/2}}{(4p_{3/2}+2p_{1/2})/6-s_{1/2}}
\end{eqnarray}
which for the experimental data shown above is of order 0.07 both
for $D$ and $B$ mesons.

\section{A dynamical spin and orbital angular momentum symmetry for the Dirac
Hamiltonian}

As it is well known, the solution of the Dirac equation with Coulomb
potential for the hydrogen atom does give spin-orbit splitting between $2p_{%
\frac{3}{2}}$ and $2p_{\frac{1}{2}}$ energy levels in agreement with
experiment. Since the one gluon exchange potential (OGE) in pQCD is
Coulomb like, one would expect [5] such a splitting in the spectrum
of hadrons, in particular heavy mesons considered here. Thus we take
$H_{q}$ in the Eq. $\eqref{2}$ as the Dirac Hamiltonian [setting
$\hbar =c=1$]
\begin{equation}
H=M_{Q}+\overrightarrow{\alpha }.\overrightarrow{p}+\beta
(m+V_{s})+V_{v} \label{5a}
\end{equation}%
where $\overrightarrow{p}=-i\overrightarrow{\nabla }$ is the
3-momentum operator, $\overrightarrow{\alpha}$ and $\beta $\ are the
Dirac matrices, $m$ is the mass of the light quark and $M_{Q}$ that
of the heavy quark $Q$. The above Hamiltonian as remarked earlier
describes the motion of light quark(antiquark) in a fixed potential
produced by the heavy antiquark(quark) as in the hydrogen atom where
electron moves in Coulomb potential provided by the nucleus(proton).
We have assumed that vector and scalar potentials are present, the
latter for the reason  to be stated shortly.

It has been observed [4,6] that if vector and scalar potentials
satisfy the relation
\begin{equation}
V_{v}(\overrightarrow{r})=V_{s}(\overrightarrow{r})+U  \label{5b}
\end{equation}%
where U is independent of the position of the light quark relative
to the heavy quark, then the Dirac Hamiltonian is invariant under a
spin symmetry (called relativistic spin symmetry)
\begin{equation}
\lbrack H,S_{i}]=0.  \label{5c}
\end{equation}%
and if potentials are spherically symmetric, then there is an
additional symmetry
\begin{equation}
\lbrack H,L_{i}]=0  \label{5d}
\end{equation}%
The generators of these symmetries are given by
\begin{equation*}
S_{i}=%
\begin{pmatrix}
s_{i} & 0 \\
0 & \tilde{s}_{i}%
\end{pmatrix}%
,L_{i}=%
\begin{pmatrix}
l_{i} & 0 \\
0 & \tilde{l}_{i}%
\end{pmatrix}%
\end{equation*}%
here $s_{i}=\frac{\sigma _{i}}{2}$ are usual spin generators,
$\sigma _{i}$
the Pauli matrices, $l_i=(r\times p)_i$ while $\widetilde{s}_{i}=U_{p}s_{i}U_{p}$ and $%
\widetilde{l}_{i}=U_{p}l_{i}U_{p},$\ with
$U_{p}=\frac{\mathbf{\sigma .p}}{2} $ as the helicity operator. Thus
even though the system may be highly relativistic, the Dirac
eigenstates can be labeled with orbital angular momentum as well as
spin, and the states with the same orbital angular
momentum are degenerate, e.g the states $n_{r}p_{\frac{1}{2}}$ and $n_{r}p_{%
\frac{3}{2}}$ are degenerate where n$_{r}$ is the radial quantum
number[4].

Thus a symmetry has been identified in the heavy-light quark system
which produces spin-orbit degeneracies independent of the details of
the potential. It may be pointed out that for the hydrogen atom,
such a symmetry would be contrary to experiment since the splitting
between energy levels $2p_{3/2} $ and $2p_{1/2}$ as predicted by
Dirac equation is in agreement with experiment which has been
regarded as a great success of Dirac equation. The levels $s_{1/2}$
and $p_{1/2}$ are still degenerate (the Lamb shift) which requires
quantum radiative corrections.

\section{Dirac equation and spin symmetry}

By writing the Dirac equation in two component Pauli form, the
symmetry discussed in the previous section can be derived in a much
more transparent way which would also help us to solve exactly the
Dirac equation if both the vector and scalar potentials are Coulomb
like. Consider the Dirac equation in covariant form in the presence
of a gauge field $V_{\mu }$
\begin{equation}
(i\gamma^{\mu }D_{\mu }-m)\Psi=0 \label{6}
\end{equation}
where $D_{\mu}=\partial _{\mu}+iV_{\mu}$ and $\gamma^{0}=\beta$,
$\gamma ^{i}=\beta \alpha ^{i}$. For our case taking the zeroth
component of $V_{\mu }$ and a scalar potential $V_s$, the above
equation become
\begin{equation}
\lbrack i\gamma ^{0}(\partial _{0}+iV_{v}(r))+i\gamma ^{i}\partial
_{i}-m-V_{s}(r)]\Psi =0.  \label{7}
\end{equation}
Multiplying on the left by
\begin{equation}
\lbrack i\gamma ^{0}(\partial _{0}+iV_{v}(r))+i\gamma ^{j}\partial
_{j}+m+V_{s}(r)]  \label{8}
\end{equation}%
we obtain, since the potentials are independent of time,
\begin{equation*}
\lbrack -\partial _{0}^{2}+2iV_{\nu }(r)\partial
_{0}+V_{v}^{2}(r)-\gamma ^{j}\gamma ^{i}(\partial _{j}\partial
_{i})+i\gamma ^{0}\gamma ^{i}[\partial _{i},V_{v}]-i\gamma
^{i}[\partial _{i},V_{s}]-(m+V_{s}(r))^{2}]\Psi =0.
\end{equation*}%
For stationary states, $\frac{\partial }{\partial t}\rightarrow iE$,
and using the fact that $\partial _{j}\partial _{i}$ is symmetric in
$j$ and $i$, we obtain%
\begin{equation}
\lbrack \nabla ^{2}+V_{v}^{2}-V_{s}^{2}-2EV_{v}-2mV_{s}+i\gamma
^{0}\gamma ^{i}[\partial _{i},V_{v}]-i\gamma ^{i}[\partial
_{i},V_{s}]+(E^{2}-m^{2})]\Psi =0,  \label{9}
\end{equation}%
where
\begin{eqnarray}
\lbrack \partial _{i},V_{v}] &=&\frac{\partial V_{v}}{\partial x^{i}}=\frac{%
\partial V_{v}}{\partial r}(\widehat{r})^{i},  \notag \\
\lbrack \partial _{j},V_{s}] &=&\frac{\partial V_{s}}{\partial x^{j}}=\frac{%
\partial V_{s}}{\partial x^{i}}(\widehat{r})^{j}.  \label{10}
\end{eqnarray}%
and the second equality holds for spherically symmetric potentials.

It is convenient to use the chiral representation of
$\gamma-$matrices,
\begin{eqnarray}
\gamma ^{0} &=&\beta =%
\begin{pmatrix}
0 & 1 \\
1 & 0%
\end{pmatrix}%
,\gamma ^{5}=i\gamma ^{0}\gamma ^{1}\gamma ^{2}\gamma ^{3}=%
\begin{pmatrix}
-1 & 0 \\
0 & 1%
\end{pmatrix}%
,  \notag \\
\alpha ^{i} &=&\beta \gamma ^{i}=%
\begin{pmatrix}
-\sigma ^{i} & 0 \\
0 & \sigma ^{i}%
\end{pmatrix}.
\label{11}
\end{eqnarray}%
Then we can write the above equation in two component matrix form%
\begin{equation}
\left( \hat{O}%
\begin{pmatrix}
1 & 0 \\
0 & 1%
\end{pmatrix}%
+i\mathbf{\sigma }.\hat{r}%
\begin{pmatrix}
-\frac{dV_{v}}{dr} & \frac{dV_{s}}{dr} \\
-\frac{dV_{s}}{dr} & \frac{dV_{v}}{dr}%
\end{pmatrix}%
\right)
\begin{pmatrix}
\Psi _{L} \\
\Psi _{R}%
\end{pmatrix}%
=0,  \label{12}
\end{equation}%
where%
\begin{equation}
\hat{O}=\nabla
^{2}+V_{v}^{2}-V_{s}^{2}-2EV_{v}-2mV_{s}+(E^{2}-m^{2}). \label{13}
\end{equation}%
The diagonalization of the matrix multiplying
$i\mathbf{\sigma}.\hat{r}$ ,
since\ $(\mathbf{\sigma }.\hat{r})^{2}=1,$ gives the eigenvalues $\eta $%
\begin{equation*}
(\frac{dV_{v}}{dr}-\eta )(-\frac{dV_{v}}{dr}-\eta )-(-(\frac{dV_{s}}{dr}%
)^{2})=0,
\end{equation*}
or
\begin{equation}
\eta =\pm \lbrack (\frac{dV_{v}}{dr})^{2}-(\frac{dV_{s}}{dr})^{2}]^{\frac{1}{%
2}}.  \label{14}
\end{equation}%
The matrix which diagonalizes this matrix does not affect the first
term in the equation $\eqref{12}$ as it is multiplied by a unit
matrix which
commutes with every matrix. Denoting the corresponding eigenfunctions by $%
\Psi _{\pm }$ which are linear combinations of $\Psi _{L}$ and $\Psi
_{R},$
we have%
\begin{equation}
\lbrack \hat{O}\pm i\eta \mathbf{\sigma }.\hat{r}]\Psi _{\pm }=0.
\label{15}
\end{equation}%
We note that the eigenvalues $\eta $ in Eq. $\eqref{14}$ vanish for
\begin{equation}
\frac{dV_{v}}{dr}=\pm \frac{dV_{s}}{dr},  \label{16}
\end{equation}%
or%
\begin{equation}
V_{v}(r)=\pm V_{s}(r)+\text{constant}.  \label{17}
\end{equation}%
Then%
\begin{equation}
\hat{O}\Psi _{\pm }=0,  \label{18}
\end{equation}%
where $\hat{O}$ is independent of spin. If plus sign is selected,
the system is said to have relativistic spin-orbital angular
momentum symmetry; for negative sign it is known as the pseudo spin
symmetry[6] which has been observed in nuclei[7]. As a result there
is no spin-orbit coupling and the results obtained in the previous
section are derived in a different and more transparent way.
Selecting the positive sign in Eqs. $\eqref{17}$ and $\eqref{18}$ we
can solve the equation
\begin{equation}
\lbrack \hat{O}\pm i\eta \mathbf{\sigma }.\hat{r}]\Psi _{\pm }=0,
\label{aa}
\end{equation}%
exactly for the energy eigenvalues as in hydrogen atom for
\begin{eqnarray}
V_{v} &=&-\frac{\alpha _{v}}{r}+U_{v},  \notag \\
V_{s} &=&-\frac{\alpha _{s}}{r}+U_{s}.  \label{19}
\end{eqnarray}%
Here $V_{v}(r)$ is the vector OGE potential with $\alpha
_{v}=\frac{4}{3}\alpha _{G} $ in which we are interested and
$V_{s}(r)$ is the confining scalar potential which might arise from
multi gluon effect. But its origin to be Coulomb like is not clear
(see, however ref [4] and references there in).

For the above case we have
\begin{equation}
\eta =\frac{(\alpha _{s}^{2}-\alpha _{v}^{2})^{\frac{1}{2}}}{r^{2}}.
\label{20}
\end{equation}%
Using the spherical polar coordinates the operator $\hat{O}$ becomes%
\begin{equation}
\hat{O}=\frac{\partial ^{2}}{\partial r^{2}}+\frac{2}{r}\frac{\partial }{%
\partial r}-\frac{L^{2}}{r^{2}}+\frac{\alpha _{v}^{2}-\alpha _{s}^{2}}{r^{2}}%
+2(E^{^{\prime }}\alpha _{v}+m^{^{\prime }}\alpha _{s})\frac{1}{r}%
+(E^{^{\prime }2}-m^{^{\prime }2}),  \label{21}
\end{equation}%
where%
\begin{eqnarray}
E^{^{\prime }} &=&E-U_{v},  \label{22} \\
m^{^{\prime }} &=&m+U_{s}.  \notag
\end{eqnarray}%
Accordingly the equation $\eqref{15}$ becomes%
\begin{equation}
\lbrack \hat{O}+i\frac{(\alpha _{v}^{2}-\alpha _{s}^{2})^{\frac{1}{2}}}{r^{2}%
}\mathbf{\sigma .r}]\psi _{+}=0.  \label{23}
\end{equation}%
The energy eigenstates can be now read of from those for the
hydrogen atom [8] by making the substitutions
\begin{eqnarray}
\alpha &\rightarrow &\alpha _{v}\frac{E^{^{\prime }}}{m^{^{\prime
}}}+\alpha
_{s},  \label{24} \\
E &\rightarrow &\frac{E^{^{\prime }2}-m^{^{\prime }2}}{2m^{^{\prime
}}}, \notag
\end{eqnarray}%
and are given by%
\begin{equation}
\frac{E^{^{\prime }2}-m^{^{\prime }2}}{2m^{^{\prime }}}=-m^{^{\prime }}\frac{%
(\alpha _{v}\frac{E^{^{\prime }}}{m^{^{\prime }}}+\alpha _{s})^{2}}{%
2(n-\delta _{j})^{2}}.  \label{25}
\end{equation}%
Solving this quadratic equation for E$^{\prime }$ and
re-substituting the values of primed quantities given in Eq.
$\eqref{22}$
\begin{equation}
E_{nj}=U_{v}+\frac{(m+U_{s})}{(n-\delta _{j})^{2}+\alpha
_{v}^{2}}[-\alpha _{v}\alpha _{s}\pm (n-\delta _{j})[(n-\delta
_{j})^{2}+\alpha _{v}^{2}-\alpha _{s}^{2}]^{\frac{1}{2}}],
\label{26}
\end{equation}%
where $\delta _{j}$ is%
\begin{equation}
\delta _{j}=(j\pm \frac{1}{2})-[(j\pm \frac{1}{2})^{2}-(\alpha
_{v}^{2}-\alpha _{s}^{2})]^{\frac{1}{2}}.  \label{27}
\end{equation}%
This is an exact result and $n$ is the principle quantum number. When $%
V_{v}=V_{s},$ so that $\alpha _{v}=\alpha _{s}$ and $U_{v}=U_{s}$
$\delta _{j}$\ becomes zero and
\begin{equation*}
E=U+(m+U)(1-\frac{2\alpha _{v}^{2}}{\alpha _{v}^{2}+n^{2}}).
\end{equation*}%
As the expression for energy is independent of $j$, there is no spin
orbit-splitting for this case.

\section{Spin-Orbit Splitting and Conclusions}

We now discuss spin-orbit splitting based on equations $\eqref{26}$
and $\eqref{27}$. For our case
$l=1$, $n=2$ and we have $1^{+}$ and $2^{+}$\ mesons corresponding to $j=%
\frac{1}{2}$ and $j=\frac{3}{2}$. Since experimentally $2^{+}$
mesons are heavier than $1^{+}$\ mesons, i.e.
$E_{2,\frac{3}{2}}>E_{2,\frac{1}{2}}$, one may conclude that $\delta
_{\frac{1}{2}}>\delta _{\frac{3}{2}}$. Then Eq.
$\eqref{27}$ implies that $(\alpha _{v}^{2}-\alpha _{s}^{2})>0$\ and further since $%
\delta _{\frac{1}{2}}$\ is real $(\alpha _{v}^{2}-\alpha _{s}^{2})<1$. For $%
V_{v}(r)$ we take OGE\ potential%
\begin{equation}
V_{v}(r)=-\frac{\alpha _{v}}{r}+U_{v}  \label{28}
\end{equation}%
\begin{figure}[tbp]
\begin{center}
\includegraphics[width=10cm]{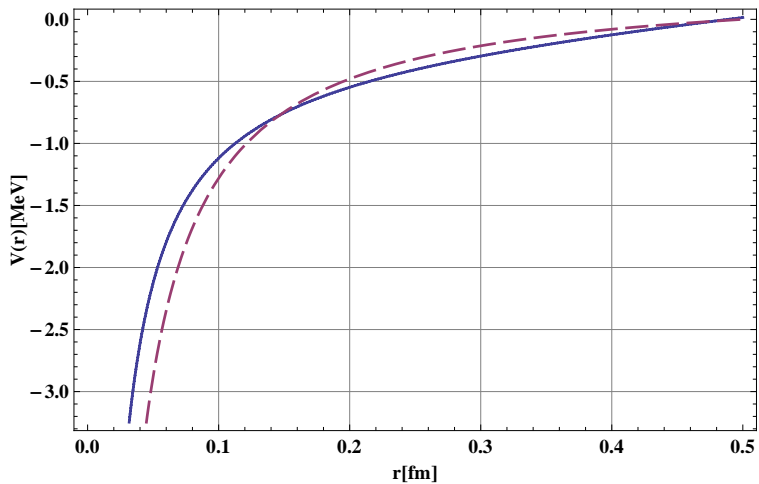}
\end{center}
\caption{Matching of OGE (Coulomb like potential, dashed) with
Cornell potential (Coulomb plus linear $r$ dependence, solid)}
\end{figure}
where $\alpha _{v}$\ is treated as an effective coupling constant.
We put $\alpha _{s}=0$\ and represent the scalar potential by a
constant $U_{s}$, the whole purpose of which is to renormalize the
light quark mass in hadron to $m+U_{s}$\ [c.f Eq. $\eqref{26}$]
where $U_{s}$\ is to be fixed by the data on spin-orbit splitting.
How good is this approach? To see this, for the charmed sector, we
compare the potential given in Eq. $\eqref{28}$ with the Cornell
potential[9]
\begin{equation}
V(r)=-\frac{K}{r}+\frac{r}{a^{2}}+C  \label{29}
\end{equation}%
with $K=0.48$, $a=2.34GeV^{-1}$ and $C=-0.25GeV$. This comparison is
shown in Fig 1 with the boundary condition ($f$ stands for fermi),
$V(0.5f)=0$ and the matching point $r=0.14f$, which gives $\alpha
_{v}=0.8$, $U_{v}=0.32GeV$. With these two conditions, which the
Cornell potential also satisfies, it is matched with the lattice QCD
potential[10] in ref. [11], showing that the Cornell potential gives
the simplest extrapolation of the lattice QCD potential. The
potential given in Eq. $\eqref{28}$ with $\alpha _v=0.8$ and
$U_r=0.32GeV$ almost give the same extrapolation as is clear from
Fig.1.

We now proceed with the numerical results. First we note that the mass of $%
\overline{Q}q$\ or $\overline{q}Q$\ meson is given by ($n=2$)
[$E_{2j}$ is given in Eq. $\eqref{26}$]
\begin{equation}
M_{2j}=M_{Q}+E_{2j}.  \label{30}
\end{equation}%
Then the mean mass of $j=\frac{1}{2}$, $j=\frac{3}{2}$ mesons is
\begin{equation}
\overline{M}=M_{Q}+\frac{1}{2}(E_{2,\frac{3}{2}}+E_{2,\frac{1}{2}%
})=M_{Q}+U_{v}+\frac{1}{2}(m+U_{s})(F_{2,\frac{3}{2}}+F_{2,\frac{1}{2}})
\label{31}
\end{equation}%
while the mass splitting is given by%
\begin{equation}
\Delta M=(E_{2,\frac{3}{2}}-E_{2,\frac{1}{2}})=(m+U_{s})(F_{2,\frac{3}{2}%
}-F_{2,\frac{1}{2}})  \label{32}
\end{equation}%
where [with $\alpha _{s}=0$]%
\begin{equation}
F_{2j}=\frac{1}{(2-\delta _{j})^{2}+\alpha _{v}^{2}}[(2-\delta
_{j})[(2-\delta _{j})^{2}+\alpha _{v}^{2}]]^{\frac{1}{2}} \label{33}
\end{equation}%
To carry out the numerical work we have to fix light quark $u$ or
$s$ mass $m$, $M_{Q}$ and $U_{s}$. It is known from the mass spectra
of $l=0$ mesons that
\begin{eqnarray}
m &=&m_{u}=330MeV  \notag \\
m_{s} &=&550MeV  \notag \\
M_{c} &=&1480MeV  \notag \\
M_{b} &=&4800MeV  \notag \\
\alpha _{v} &=&\frac{4}{3}\alpha _{G},
\end{eqnarray}%
where $\alpha _{v}$ and $U_{v}$ have already been fixed for the
charmed sector. As the QCD coupling $\alpha _{v}$\ is energy
dependent, decreasing with increasing energy, one would expect
\begin{equation*}
\alpha _{v}(M_{B_{s}})<\alpha _{v}(M_{B})<\alpha _{v}(M_{D}),
\end{equation*}%
we will take this into consideration.

It is clear from Eqs. $\eqref{31}$, $\eqref{32}$ that once we have
calculated $F_{2j},(m+U_{s})$ can be fixed from Eq. $\eqref{31}$.
Our numerical results are summarized in Tables I and II.

\begin{table}[htb]
\caption{Experimental data; all masses in MeV}
\begin{tabular}{cccc}
\hline \hline $\text{Mesons}$ & $\hspace{1cm}\text{Mass }M$ &
$\hspace{1cm}\text{Mean Mass }\overline{M}$ &
$\hspace{1cm}\text{Mass Splitting }\Delta M$ \\
\hline $D_{1}^{\ast }(1^{+})$ & $\hspace{1cm}2422$ &
$\hspace{1cm}2442$ & $\hspace{1cm}40$ \\
\hline $D_{2}^{\ast }(2^{+})$ & $\hspace{1cm}2462$ &
$\hspace{1cm}$ & $\hspace{1cm}$ \\
\hline $B_{1}^{\ast }(1^{+})$ & $\hspace{1cm}5723$ &
$\hspace{1cm}5733$ & $\hspace{1cm}21$ \\
\hline $B_{2}^{\ast }(2^{+})$ & $\hspace{1cm}5744$ &
$\hspace{1cm}$ & $\hspace{1cm}$ \\
\hline $B_{s_{1}}^{\ast }(1^{+})$ & $\hspace{1cm}5829$ &
$\hspace{1cm}5834$ & $\hspace{1cm}11$ \\
\hline $B_{s_{2}}^{\ast }(2^{+})$ & $\hspace{1cm}5840$ &
$\hspace{1cm}$ & $\hspace{1cm}$ \\
\hline \hline
\end{tabular}
\end{table}

\begin{table}[htb]
\caption{Predicted mass splittings in MeV}
\begin{tabular}{cccccccc}
\hline \hline $\text{Flavor}$ & $\hspace{1cm}\alpha _{v}$ &
$\hspace{1cm}U_v$ & $\hspace{1cm}F_{\frac{3}{2}}-F_{\frac{1}{2}}$ &
$\hspace{1cm}(F_{\frac{3}{2}}+F_{\frac{1}{2}})/2$ &
$\hspace{1cm}m+U_s$
& $\hspace{1cm}\Delta M\text{(our)}$ & $\hspace{1cm}\Delta M\text{(expt)}$ \\
\hline $D$ & $\hspace{1cm}0.8$ & $\hspace{1cm}320$ &
$\hspace{1cm}0.029$ & $\hspace{1cm}0.901$ & $\hspace{1cm}712$ &
$\hspace{1cm}21$ & $\hspace{1cm}40$ \\
\hline $B$ & $\hspace{1cm}0.75$ & $\hspace{1cm}300$ &
$\hspace{1cm}0.019$ & $\hspace{1cm}0.917$ & $\hspace{1cm}690$ &
$\hspace{1cm}13$ & $\hspace{1cm}21$ \\
\hline $B_s$ & $\hspace{1cm}0.65$ & $\hspace{1cm}260$ &
$\hspace{1cm}0.009$ & $\hspace{1cm}0.938$ & $\hspace{1cm}814$ &
$\hspace{1cm}7.3$ & $\hspace{1cm}11$ \\
\hline \hline
\end{tabular}
\end{table}

To conclude we see that the spin-orbit splittings, obtained from the
Dirac equation with OGE potential and scalar confining potential to
be a constant so that its role is to enhance the mass of light quark
in a hadron, are small and qualitatively explain the data within a
factor of about 2 (for charm) and 1.5 (for bottom). The values used
for $\alpha _v$ for the OGE potential are some what larger from
those obtained from the asymptotic freedom of QCD. However, in
potential models $\alpha _v$ is usually treated as a
phenomenological parameter. From Table 2, we see that $U_s$ for the
scalar confining potential is almost independent of flavor. Another
important conclusion one can draw is that as in hydrogen atom, the
meson spectroscopy does not show relativistic spin and orbital
angular momentum symmetry, since the splittings are small anyway.

\section*{Acknowledgments}
One of the author (SS) would like to acknowledge the support of
Higher Education Commission (HEC), Pakistan, under the indigenous
scholarship.

\end{document}